\documentstyle[prl,aps,twocolumn,graphicx,floats]{revtex}
\newcommand{\eq}[1]{Eq. (\ref{#1})}

\begin{document}

\title{Resistive Anomalies at Ferromagnetic Transitions Revisited: the
case of SrRuO$_3$ }
\author{R. Roussev and A. J. Millis}

\address{Center for Materials Theory and Department of Physics and Astronomy, Rutgers University, 136 Frelinghuysen Road, Piscataway NJ 08854-8019}

\maketitle

It is generally believed \cite{Fisher68,Alexander76} that near a
ferromagnetic phase transition the resistivity, $\rho$, exhibits an
energy-like singularity so ${\rm d} \rho / {\rm d} T \sim
|t|^{-\alpha}$ with $\alpha \approx 0.1$ the specific heat exponent and
$t=(T-T_c)/T_c$.  In a recent Letter \cite{Klein96}, Klein {\em et al.}
claimed that in SrRuO$_3$, a strongly correlated ``bad metal'', the
resistive anomaly at $T_c$ is anomalous: at $T>T_c$, ${\rm d} \rho /
{\rm d} T \sim t^{-x}$ with $x \approx 0.9$ while for $T<T_c$ ${\rm d}
\rho / {\rm d} T $ cannot be fit by any power law.  Klein {\em et al.}
obtained exponents by plotting $\ln |{\rm d} \rho / {\rm d} T - S_0|$
vs $\ln T/T_c$ ($S_0$ is a $T$-independent background), a procedure
shown to be unreliable in many cases \cite{Ahlers80,CAS}.

\begin{figure}
\includegraphics[scale=0.5]{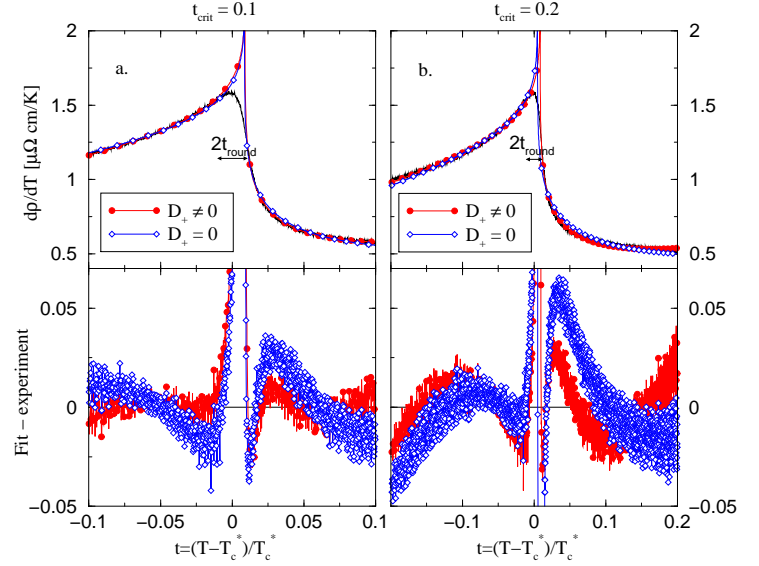}
\caption{\label{figure} Experimental data \protect\cite{Klein96} and
Heisenberg fits with \eq{bestfit} (upper graphs) for $t_{\rm
round}=0.01$.  The lower graphs show the difference between
\protect\eq{bestfit} and the data.  $T_c^*=150$ K.}
\end{figure}

We show here that the data are consistent with conventional theory,
\cite{Fisher68,Alexander76} which predicts
\begin{equation}
{\rm d}\rho/{\rm d}T =
(A_{\pm} / \alpha)|t|^{-\alpha}(1+D_{\pm}|t|^\theta)+S(t),
\label{bestfit}
\end{equation}
where $A_\pm$ and $D_\pm$ are the amplitudes for the leading
singularity and correction to scaling, $\alpha$ and $\theta$ are the
specific heat and correction to scaling exponents and $+(-)$ refer to
$t>0(t<0)$. $S(t)$ is a smooth function of $t$ which we approximate by
$S(t) = S_0 + S_1 t$.  \eq{bestfit} applies within a critical region
$|t|<t_{\rm crit}$.  The fits to $\chi$ in \cite{Klein96} imply $t_{\rm
crit}\sim 0.13$; fitting $|t|>t_{\rm crit}$ requires a crossover theory
\cite{CAS}. Due to rounding of the data very near $T_c$ points with
$|t|<t_{\rm round}$ have been excluded.  Following \cite{Ahlers76} we
allowed for 1\% variations in $T_c$ about the nominal value
$T_c^*=150$K.  The fits depend crucially on $t_{\rm crit}$, $t_{\rm
round}$ and $t_c$.  Fig. \ref{figure} shows the data along with fits to
\eq{bestfit} using Heisenberg exponents $\alpha_H=-0.1$ and
$\theta_H=0.55$, and universal amplitude ratios $(A_+/A_-)_H=1.52$ and
$(D_+/D_-)_H=1.4$ \cite{Aharony} with and without corrections to
scaling for $t_{\rm round} = 0.01$ and parameters given in Table
\ref{params}.
Using Ising parameters in \eq{bestfit} produces
slightly worse fits (not shown). The curve {\em with} corrections to
scaling (i.e. $D_+ \ne 0$) and $t_{\rm crit} = 0.1$ is our best fit
with accepted universal amplitude ratios; varying these gives exact
fits over the range $t \in (-0.2,0.5)$.  The results of \cite{Klein96}
imply that the $t>0$ behavior is also consistent with $\alpha \approx
0.9$, but we believe the consistency with conventional theory renders
alternative interpretations implausible.

One of us \cite{Millis95} had previously suggested that in systems with
very strong carrier-spin interactions, a term in $\rho(T)$ proportional
to the square of the magnetization, ${\mathcal M} ^2$, could arise at
$T<T_c$.  This suggestion was based on an erroneous interpretation of a
spherical model calculation and is here withdrawn. Briefly, any
scattering process contributing to $\rho$ involves a combination of
spin operators at nearby points \cite{Fisher68} and in a ferromagnet
these can only involve the specific heat exponent. In the spherical
model the specific heat and ${\rm d} {\mathcal M}^2 / {\rm d} T$ have
identical behavior.

We thank L. Klein for data, M. E. Fisher, J. Ye and J. S. Dodge for
helpful conversations, J. S. Dodge for pointing out an error in our
previous analysis, and NSF - DMR - 9707701 and the Johns Hopkins
MRSEC.

\begin{table}[!hbp]
\begin{tabular}{ccccc} 
$t_{\rm crit}$ &  0.1 & 0.1 & 0.2 & 0.2 \\
\hline
$A_+$	& 0.3820 & 0.3331 & 0.3484 & 0.3052 \\
$D_+$	& -0.256 & 0.0 & -0.171 & 0.0 \\
$S_0$	& 3.238 & 3.043 & 3.071 & 2.886 \\
$S_1$	& 1.556 & 1.316 & 1.126 & 1.052 \\
$t_c$	& -0.0090 & -0.0070 & -0.0086 & -0.0048 
\end{tabular}
\caption{\label{params} Fitting parameters for \eq{bestfit} with
$t\rightarrow t+t_c$.}
\end{table}

\end{document}